# Energetic Electron Precipitation Occurrence Rates Determined Using the Syowa East SuperDARN Radar

**Emma C. Bland[1,2], Noora Partamies[1,2], Erkka Heino[2,3], Akira Sessai Yukimatu[4,5], and Hiroshi Miyaoka[4,5]**

[1]Birkeland Centre for Space Science, Bergen, Norway, [2]Department of Arctic Geophysics, University Centre in Svalbard, Longyearbyen, Norway, [3]Department of Physics and Technology, University of Tromsø, Tromsø, Norway, [4]National Institute of Polar Research, Tokyo, Japan, [5]The Graduate University for Advanced Studies (SOKENDAI), Hayama, Japan

**Key Points:**

- *D* region ionization due to energetic electron precipitation can be detected using a high-frequency radar at Syowa station, Antarctica
- Event detection method is validated using all-sky camera observations of pulsating aurorae
- Energetic electron precipitation occurrence rates peak in the morning sector and in the winter

**Correspondence to:**
E. C. Bland,
emmab@unis.no

## Abstract

We demonstrate that the Super Dual Auroral Radar Network (SuperDARN) radar at Syowa station, Antarctica, can be used to detect high frequency radio wave attenuation in the *D* region ionosphere during energetic electron precipitation (EEP) events. EEP-related attenuation is identified in the radar data as a sudden reduction in the backscatter power and background noise parameters. We focus initially on EEP associated with pulsating aurora and use images from a colocated all-sky camera as a validation data set for the radar-based EEP event detection method. Our results show that high-frequency attenuation that commences during periods of optical pulsating aurora typically continues for 2–4 hr after the camera stops imaging at dawn. We then use the radar data to determine EEP occurrence rates as a function of magnetic local time (MLT) using a database of 555 events detected in 2011. EEP occurrence rates are highest in the early morning sector and lowest at around 15:00–18:00 MLT. The postmidnight and morning sector occurrence rates exhibit significant seasonal variations, reaching approximately 50% in the winter and 15% in the summer, whereas no seasonal variations were observed in other MLT sectors. The mean event lifetime determined from the radar data was 2.25 hr, and 10% of events had lifetimes exceeding 5 hr.

## 1. Introduction

Energetic electron precipitation (EEP, 10–1,000 keV) is known to cause strong ionization of the mesosphere and lower thermosphere region, which in turn impacts the composition and dynamics of the Earth's atmosphere. One notable effect of this ionization is the production of odd-hydrogen ($HO_X = H + OH + HO_2$) and odd nitrogen ($NO_X = N + NO + NO_2$) species, which are catalysts for ozone depletion (Sinnhuber et al., 2012, and references therein). EEP events have been shown to cause mesospheric ozone depletion of tens of percent (Andersson et al., 2014; Daae et al., 2012; Seppälä et al., 2015; Turunen et al., 2016). In the winter, $NO_X$ can also be transported downward by the polar vortex and cause further ozone depletion in the polar stratosphere (Daae et al., 2012; Hendrickx et al., 2018; Randall et al., 2009; Smith-Johnsen et al., 2018). Due to the high occurrence rates of EEP, it is thought that EEP may have a cumulative effect on the atmosphere and potentially impact polar climate variability (Andersson et al., 2014; Rozanov et al., 2012; Seppälä et al., 2014, 2015).

The atmospheric effects of EEP have been studied using models such as the *Whole Atmosphere Community Climate Model* (Marsh et al., 2007) and the *Sodankylä Ion-Neutral Chemistry* model (Turunen et al., 2009; Verronen et al., 2005). These models have been used to study the impacts of individual EEP events (e.g., Smith-Johnsen et al., 2018; Turunen et al., 2016) and also to investigate possible impacts of EEP on climate variability (e.g., Seppälä et al., 2014). The EEP input for atmospheric models is usually parameterized using proxies for geomagnetic activity such as the *Ap* and *Kp* indices. For example, van de Kamp et al. (2016, 2018) provide *Ap* parametrizations based on 15 years of particle flux measurements from the Polar Orbiting Environmental Satellites. These parameterizations provide a good estimate of the EEP energy input during strong magnetic activity (geomagnetic storms) but may not necessarily capture the energy input associated with other types of EEP events such as local substorm activity and pulsating aurorae (PsA; Beharrell et al., 2015; Partamies et al., 2017). In particular, PsA are often observed in the substorm recovery phase and may continue for several hours after geomagnetic activity levels have recovered (Partamies et al., 2017). Therefore, the standard geomagnetic index parametrizations are likely to underestimate the total EEP energy input to the middle atmosphere.

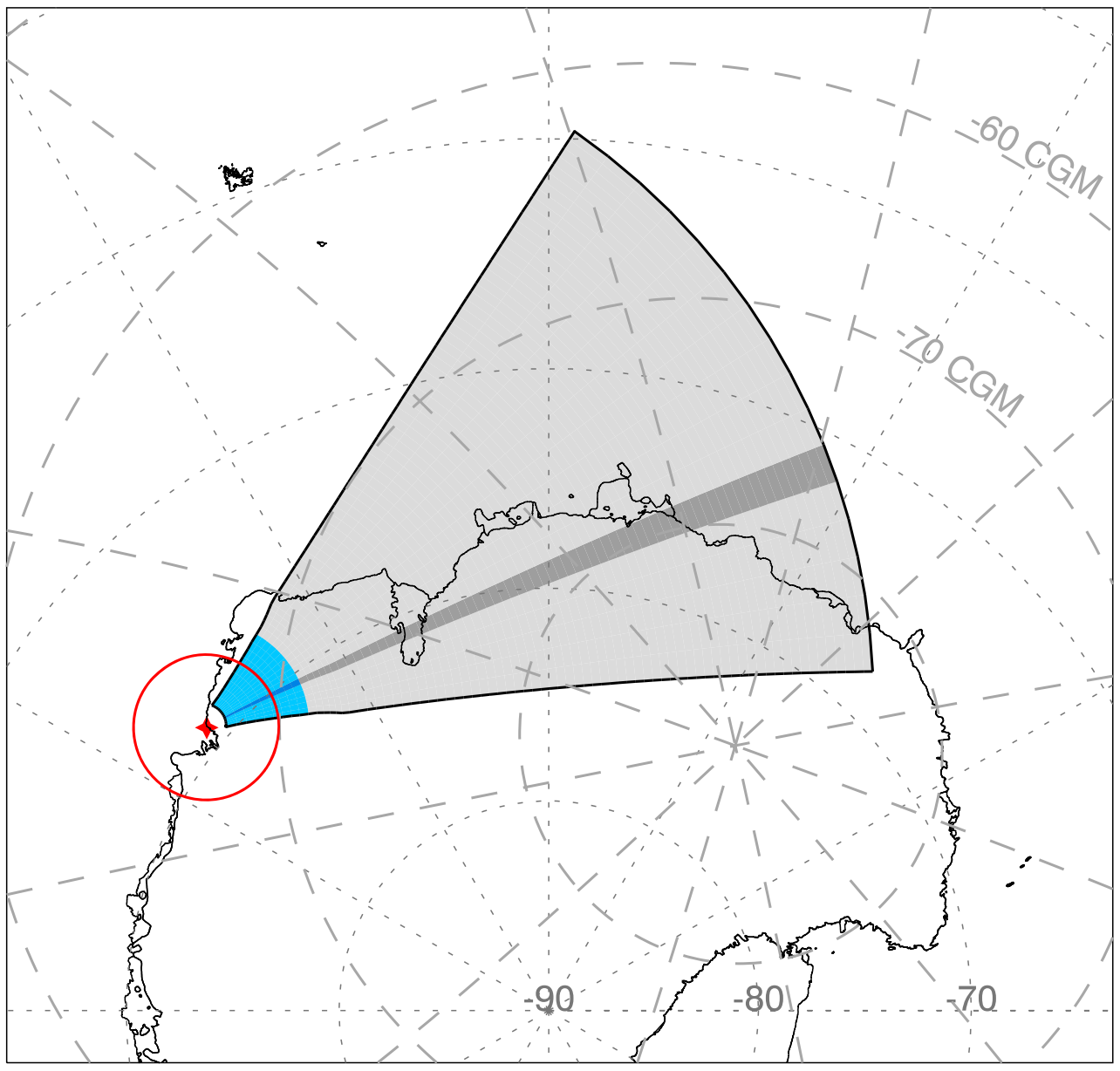

**Figure 1.** Fields of view of the all-sky color digital camera at Syowa station (red circle, projected to 100 km) and the SENSU Syowa East SuperDARN radar. The position of Beam 5 in the radar field of view is indicated by darker shading. Dotted (dashed) lines indicate geographic (geomagnetic) coordinates. SuperDARN = Super Dual Auroral Radar Network; SENSU = Syowa South and East HF Radars of NIPR for SuperDARN.

To obtain more realistic descriptions of EEP forcing, it is necessary to characterize the energy input and occurrence rates of all EEP event types, especially weaker EEP events such as PsA, which are not well described by geomagnetic indices. Spacecraft observations of EEP fluxes from, for example, the Polar Orbiting Environmental Satellites provide good estimates of the particle energy deposition for precipitating electrons (e.g., Nesse Tyssøy et al., 2019). These measurements, however, are limited in spatial, temporal, and pitch angle coverage. Ground-based instrumentation, especially riometer networks, provide multipoint observations of EEP events with better temporal coverage. Riometers are sensitive to excess ionization in the *D* region ionosphere, which leads to attenuation of the cosmic radio noise through collisions with the neutral atmospheric constituents. This method is most sensitive to precipitating electrons with energies up to the order 10–1,000 keV, which deposit their energy at around 70- to 100-km altitude (Fang et al., 2008). Statistical studies of cosmic noise absorption (CNA) from riometer networks indicate strong magnetic local time (MLT) variations in auroral zone EEP (e.g., Basler, 1963; Foppiano & Bradley, 1985; Hargreaves & Cowley, 1967a, 1967b; Kavanagh et al., 2004, 2012). These studies report a daily maximum CNA in the morning sector and a minimum near dusk. Strong seasonal variations in morning sector CNA have also been identified, with a maximum occurring in the winter (e.g., Basler, 1963; Hargreaves & Cowley, 1967b; Kavanagh et al., 2012). Seasonal differences in CNA are less pronounced on the nightside (e.g., Foppiano & Bradley, 1985).

In addition to riometer observations, several recent studies have demonstrated that the Super Dual Auroral Radar Network (SuperDARN) can be used to observe radio wave attenuation in the ionospheric *D* region using an approach similar to riometry (Berngardt et al., 2018; Bland et al., 2018; Chakraborty et al., 2018). These studies focused on detecting high-frequency (HF) radio attenuation arising from solar flares and energetic proton precipitation. In this study we show that SuperDARN radars can also detect HF radio attenuation caused by EEP, including the low particle flux PsA events. SuperDARN consists of more than 30 radars, and the data set includes over 20 years of observations, providing an opportunity to determine the statistical properties of EEP events to complement the results of earlier riometer studies. To demonstrate this capability, we use the Syowa East radar of the *Syowa South and East HF Radars of NIPR for SuperDARN* (SENSU), which observes the ionospheric region near the equatorward edge of the Southern Hemisphere auroral oval. We focus initially on detecting PsA, which can be validated using data from an all-sky camera, which is colocated with the radar. Since the radar can readily detect PsA-related attenuation, it follows that the radar can also detect attenuation caused by other types of EEP events involving higher electron number fluxes such as substorms. We then use the radar data set to determine realistic EEP occurrence rates as a function of MLT and also for different seasons.

## 2. Instrumentation

### 2.1. SENSU Syowa East SuperDARN Radar

The SENSU Syowa East (SYE) radar is part of the SuperDARN, a global network of HF coherent scatter radars designed for studying ionospheric plasma convection from middle to polar latitudes (Chisham et al., 2007; Greenwald et al., 1995; Lester, 2013). The radars detect backscatter primarily from field-aligned electron density irregularities in the *E* and *F* region ionospheres. Backscatter is also commonly detected from the ground following total internal reflection by the ionosphere and also from meteor plasma trails at around 90- to 100-km altitude (Chisham & Freeman, 2013; Hall et al., 1997; Yukimatu & Tsutsumi, 2002). The transmitted frequency can be adjusted to suit the ionospheric propagation conditions, in particular to maximize the occurrence of ionospheric backscatter. Most observations by the SYE radar are performed at 10–11 MHz.

The location and field of view (FOV) of the SYE radar are shown in Figure 1. Dotted lines indicate geographic coordinates, and dashed lines indicate altitude adjusted corrected geomagnetic coordinates (Baker & Wing, 1989). The radar is located at 69.00°S, 39.58°E geographic (altitude adjusted corrected geomagnetic: 66.5°S, 72.2°E), and the FOV covers a 52° azimuthal sector, which extends to approximately 3,500 km in range. The blue shaded region in Figure 1 indicates the portion of the FOV from which half-hop backscatter from the *D* region/lower *E* region ionosphere might be observed. Backscatter from longer ranges (gray shaded region) is expected to come from *F* region irregularities or from the ground/sea following reflection in the ionosphere (Chisham et al., 2008). The standard range gate resolution is 45 km, but the SYE radar also operates regularly with 30- or 15-km range resolution.

The SYE radar operates with 16 azimuthal beams (numbered 0–15), with the boresight oriented 106.5° (143°) from geographic (geomagnetic) north. The position of Beam 5, which will be used for illustrative purposes in section 3, is indicated by darker shading Figure 1. Each beam is narrow in azimuth (~3.24°) but has a wide vertical extent (e.g., Milan, Jones, et al., 1997). In the standard operational mode, beams are sampled consecutively with a ~3-s dwell time per beam. Thus, the total FOV is scanned every 60 s.

The standard data products of SuperDARN are power (signal-to-noise ratio), Doppler velocity, and spectral width of the received backscatter. However, in this study we make use of two lower-level data products: (1) the raw echo power (also called *lag zero power*) and (2) the background noise level at the radar operating frequency. These parameters will be used to identify evidence of radio wave attenuation caused by EEP. The raw echo power is the power of the unprocessed autocorrelation function of the backscattered signal at lag zero, measured in analog-to-digital units. This parameter is measured in all 75 range gates along the beam. High values of echo power that are grouped together in neighboring range gates and persist for tens of minutes to several hours are evidence of coherent backscatter. Low echo power in a particular range gate occurs when there are no scattering targets present at that range or when the ionosphere does not support radio propagation to that range and back to the radar. Low values of echo power observed simultaneously in all range gates indicate strong attenuation of the transmitted radio waves in the *D* region ionosphere (Bland et al., 2018).

The background noise parameter is estimated as the average of the 10 lowest values of the raw echo power, which represents the power level in range gates for which no backscatter was detected. This method provides one background noise value each time a particular beam is sampled. At ~10 MHz, the main contributor to the background noise is atmospheric noise arising from global lightning activity (Headrick & Anderson, 2008). As a result, SuperDARN background noise measurements are essentially a function of solar time but exhibit day-to-day variability (Berngardt et al., 2018; Bland et al., 2018). Very high background noise levels indicate external radio interference, most likely from man-made sources close to the radar site. Very low noise levels indicate increased radio attenuation in the *D* region ionosphere, especially when accompanied by backscatter loss (no grouped populations of enhanced echo power). This combination of reduced echo power and reduced background noise is the signature used in this study to identify radio attenuation during EEP events.

### 2.2. All-Sky Color Digital Camera

For validation purposes, we use optical observations from an all-sky color digital camera, which is colocated with the radar at Syowa station. The camera FOV projected to 100-km height is shown by the red circle in Figure 1. This overlaps with the near-range portion of the SYE radar FOV.

The imaging season at Syowa station lasts from April to October. The camera is programmed to capture images automatically whenever the Sun is more than 12° below the horizon. Images are recorded with a 6-s cadence and an exposure time of 4 s. This low sampling rate and long exposure time are insufficient to capture the temporal behavior of individual auroral pulsations, but events can be readily identified in daily keogram (quick-look) plots as patchy auroral displays (e.g., Jones et al., 2013; Partamies et al., 2017; Yang et al., 2017). Keograms are constructed by taking a slice of pixels through an all-sky image from magnetic north to south and then assembling the slices sequentially along a time axis.

## 3. HF Radar Signatures of EEP

First, we illustrate and characterize the typical behavior of the echo power and background noise parameters measured by SYE during PsA events and then show that the same features also occur for other EEP events. Figure 2 shows observations from the all-sky camera and the radar for the 12-hr interval commencing at 22:00 UT on 13 June 2015. Figure 2a shows the keogram for the geomagnetic north-south direction of the camera FOV. Two PsA events, identified visually from the keogram data, are indicated by the red square brackets above the plot. Figure 2b shows the background noise measured on each radar beam. These data have been binned into 5-min time intervals, averaged separately in each bin, and then smoothed using a two-dimensional boxcar filter of size $2 \times 2$ bins to remove spikes caused by radio interference. Following the format of the keogram, the most poleward radar beam (Beam 0) is displayed at the top of the plot. Figure 2c shows the echo power as a function of slant range for Beam 5.

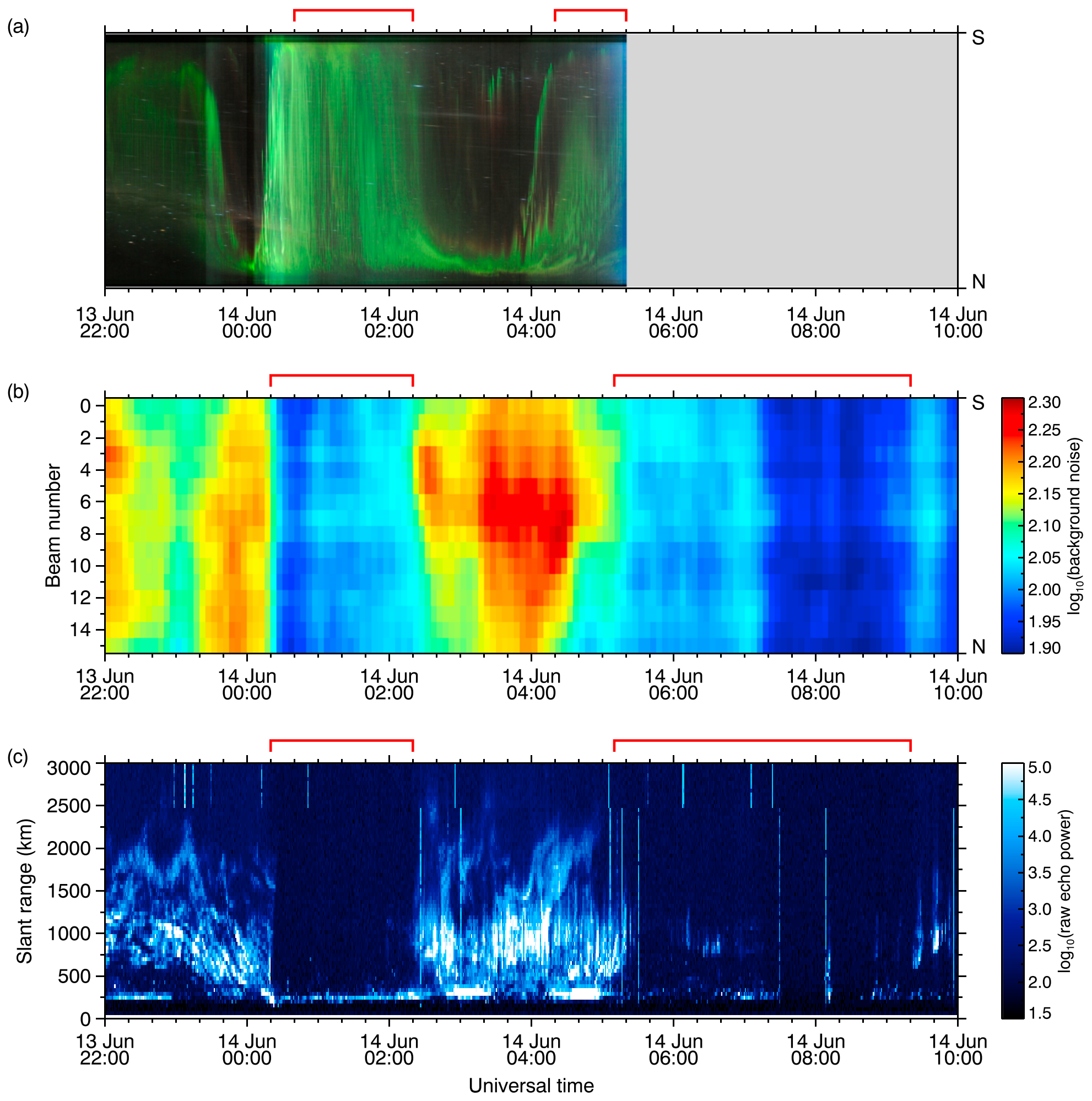

**Figure 2.** Quick-look plot used for pulsating aurora event identification for the time interval 13–14 June 2015 (22:00–10:00 UT). (a) The keogram from the all-sky camera, (b) the background noise measured along all radar beams, and (c) the echo power measured on Beam 5. The magnetic north (N) and south (S) directions are indicated on the right vertical axis for the keogram and background noise plots. Red brackets above each panel indicate pulsating aurora events.

The first PsA event identified in the keogram occurs from 00:40–02:20 UT. During this time, the radar detects a significant reduction in both the noise and echo power parameters. This feature is indicated by the square brackets above Figures 2b and 2c. We interpret these changes as evidence of enhanced radio attenuation caused by increased ionization of the *D* region during the PsA event. The decreased background noise results from attenuation of the ambient radio noise at 10 MHz, and the decreased echo power results from attenuation of the radio waves transmitted by the radar. Since the background noise and echo power respond to the presence and absence of optical PsA almost instantaneously, the radar response to PsA is easy to distinguish from the more gradual diurnal variations in the background noise. Note that the event onset time determined from the radar data is 20 min earlier than the onset time determined from the optical data. This is most likely due to electron precipitation associated with the auroral activation at 00:20–00:40 UT, which is not related to PsA.

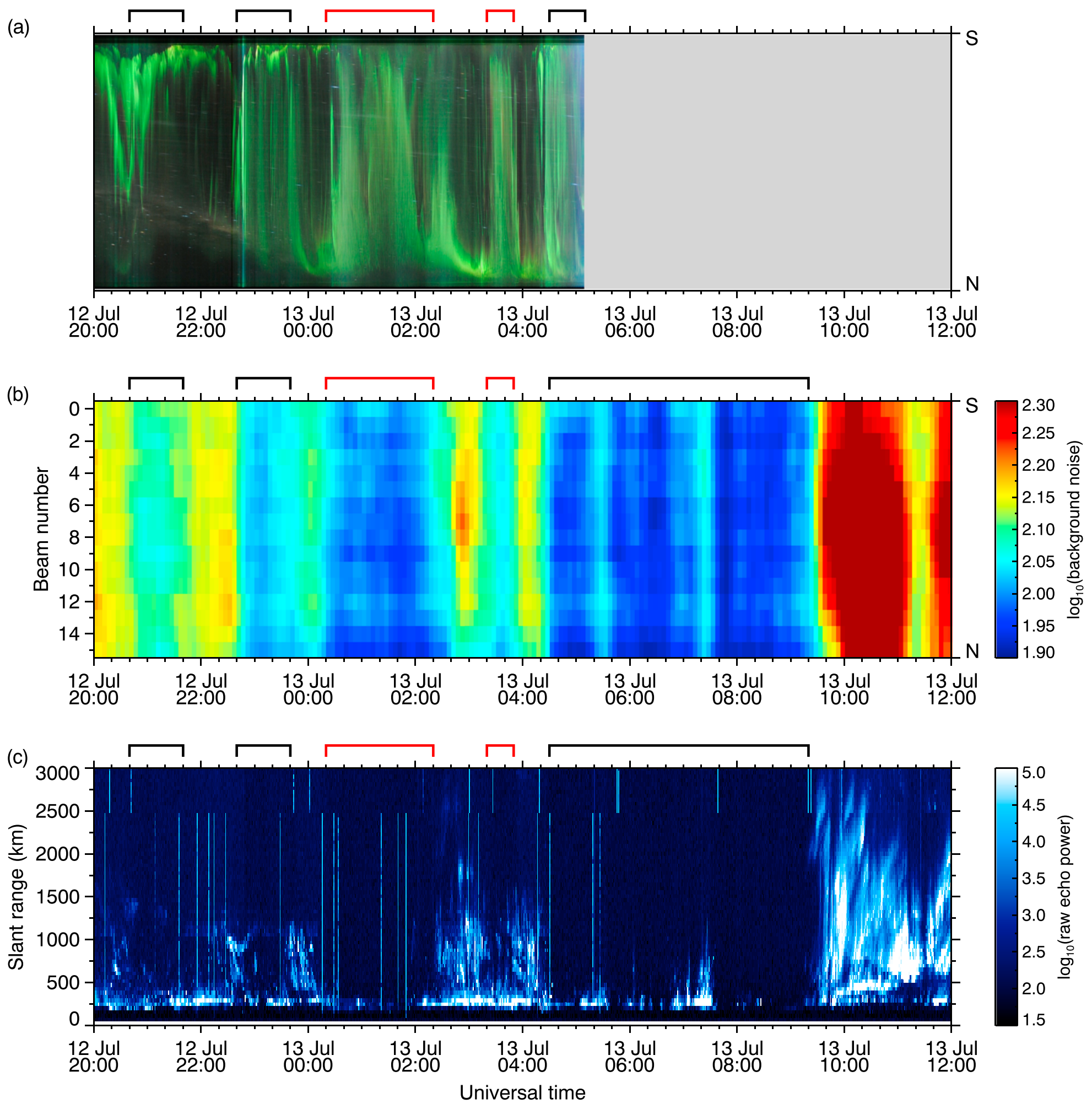

**Figure 3.** Same as Figure 2 but for the time interval 12–13 July 2015 (20:00–12:00 UT). Red brackets above each panel indicate pulsating aurora events, and black brackets indicate other auroral activations.

At 02:20 UT, the PsA shifts equatorward and is visible only at the northern edge of the camera FOV. We define this time as the end of the PsA event. The aurora remains in this portion of the camera FOV from 02:20–04:00 UT, during which time there is a significant increase in the background noise and echo power measured by the radar. The increase in both of these parameters indicates that the radio attenuation associated with the PsA has ended. Notice that the background noise increase occurs more gradually on the most equatorward beams, where optical aurorae are still present in the keogram.

A second PsA event becomes evident in the camera data at 04:20 UT. This is accompanied by a reduction in the background noise, which is first detected by the equatorward radar beams where the intensity of the optical PsA is greatest. There is also a decrease in the echo power, but this lags behind the optical observations by approximately 45 min. This might be due to a lower flux of energetic electrons into the *D* region, resulting in less radio attenuation compared to the previous PsA event. Also, since the radar beams overlap with only the poleward part of the camera FOV (Figure 1), the radar is not expected to be sensitive to *D* region enhancements, which are present only in the equatorward part of the camera FOV. This demonstrates the limited spatial extent of the attenuating region during the PsA event and the sensitivity of the radar to the location of the optical aurora. Based on the keogram data, the second PsA event in Figure 2 was still in progress when the camera stopped imaging at 05:20 UT. We have used the radar observations to estimate the true event end time by identifying a recovery in both the background noise and echo power. In this example, the recovery occurs at 09:20 UT, which is 4 hr after imaging ends. Note that without optical data to confirm the presence of PsA, it is not possible to determine if the attenuation observed by the radar after 05:20 UT is due to PsA or another type of EEP.

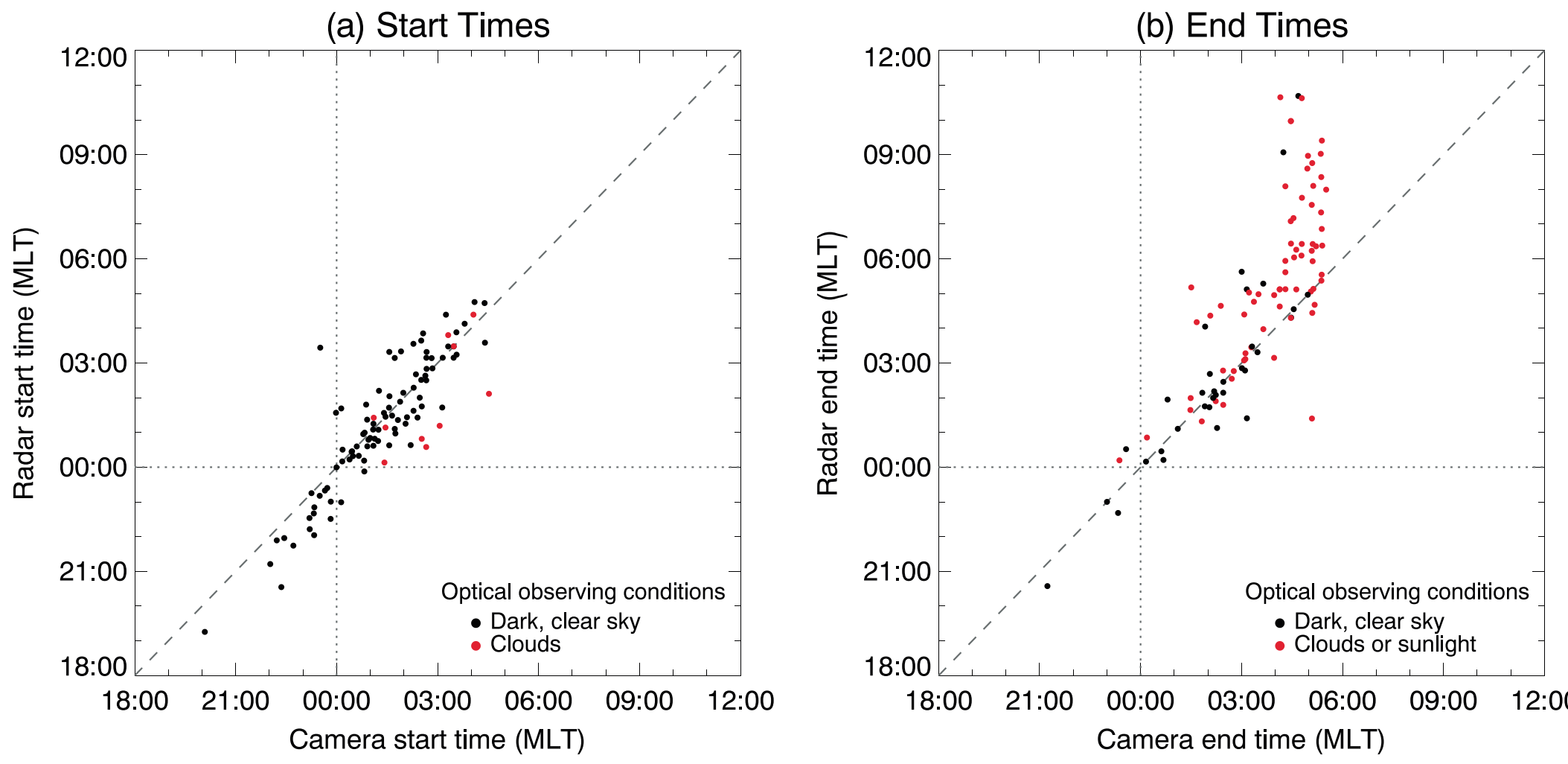

**Figure 4.** Scatter plots of (a) event start times and (b) event end times, as determined from the camera and radar data. Red symbols are used whenever the true start/end time could not be determined from the camera data due to cloud cover or daylight. MLT = magnetic local time.

Quick-look plots for another interval of auroral activity are shown in Figure 3. The 16-hr time interval shown commences at 20:00 UT on 12 July 2015. This time interval contains examples of pulsating and also non-PsA, indicated by red and black brackets, respectively. The event at 00:20–02:20 UT is mostly PsA, whereas the event at 03:20–03:50 UT contains a mixture of PsA and other aurora. The remaining three events contain no identifiable pulsating features. This classification was performed by examining the original all-sky camera images to better distinguish between PsA and non-PsA features (not shown). All five events in Figure 3a are accompanied by reduced background noise and echo loss, which closely follow the optical aurora. The final event, which commenced at 04:30 UT, is shown to continue more than 4 hr after the camera stopped imaging. During this period, there are two brief periods in which small amounts of backscatter are observed in the near-range FOV and the background noise increases slightly. These observations indicate lower levels of attenuation and could be interpreted as the end of an EEP event. However, large amounts of backscatter are not detected until the noise level increases significantly at 09:20 UT, indicating a sustained enhancement of the *D* region ionosphere from 04:30–09:20 UT. Therefore, we choose to classify this entire period as a single EEP event. This very short range backscatter is also present during the first event shown in Figure 2 and may result from the enhancement of *E* region irregularities by the precipitating electrons and low aspect angle sensitivity of the backscatter (Milan et al., 2004). Therefore, in determining the time limits of an EEP event we look for the presence or absence of *F* region backscatter in particular. Further discussion on this short-range backscatter is provided in section 6.

The results shown in Figure 3 demonstrate that the SYE radar observes HF attenuation not only during PsA events but also during other types of EEP events involving higher electron number fluxes into the *D* region. We note that there is no clear feature in the radar observations, which would allow us to distinguish between PsA and non-PsA periods based on the radar data alone. However, this distinction is unnecessary in this study since we seek to determine occurrence rates of EEP events of any type. The PsA events are used only as a "threshold" type of event to demonstrate that the radar is capable of detecting attenuation due to very low electron fluxes and which can be validated using the camera data when available. In the following statistical survey, we distinguish between events for which the keogram data have been viewed to confirm the presence of PsA and attenuation events identified using only the radar data, which we attribute to all types of EEP. For optically confirmed PsA, we determine the durations and MLT distribution of PsA-related EEP using the radar-determined event lifetimes, which are independent of the optical observing conditions (section 4). For events identified using only the radar data, we determine occurrence rates of all EEP-related attenuation events (section 5).

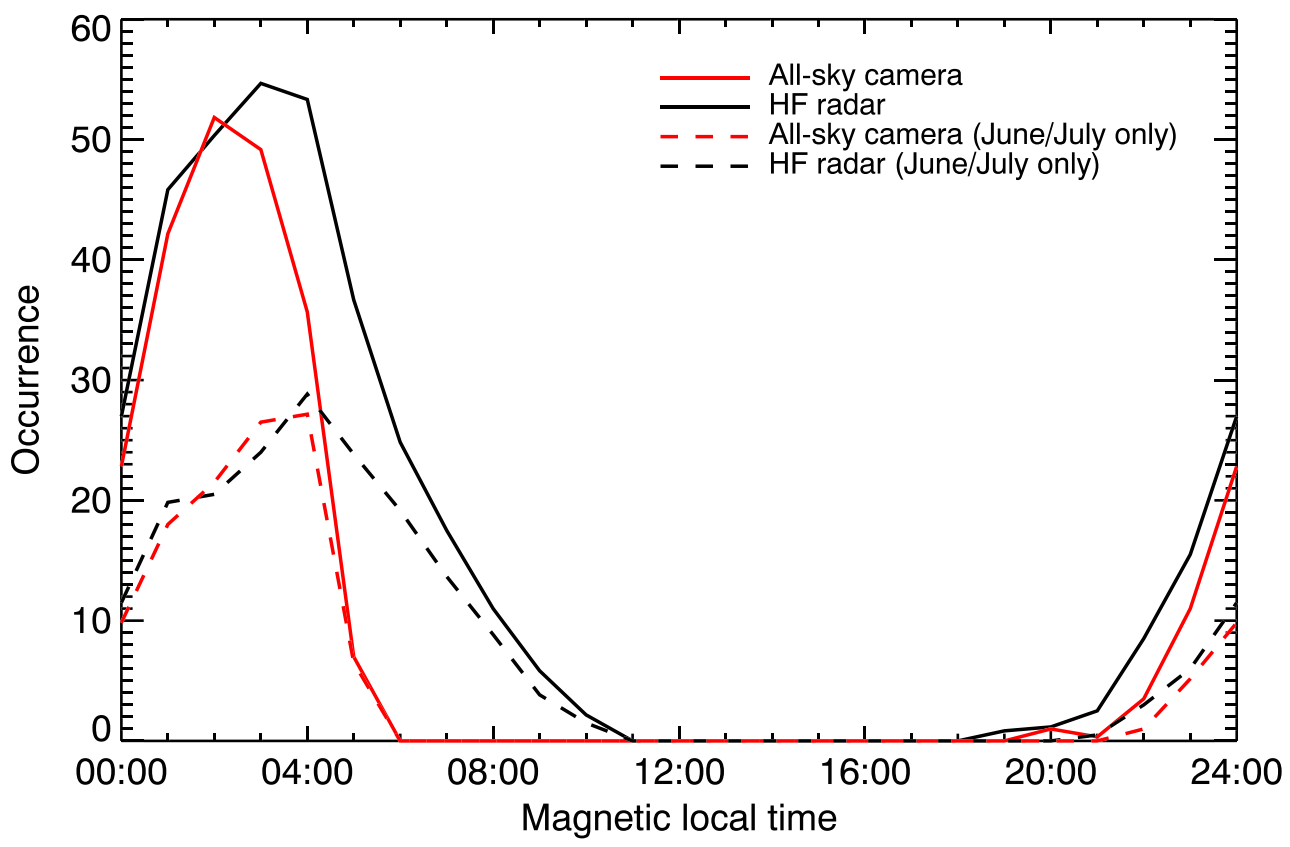

**Figure 5.** Pulsating aurora event occurrence as a function of magnetic local time, determined using the camera data (red line) and radar data (black line). HF = high-frequency.

## 4. Statistics for Optically Confirmed PsA

The results presented in the previous section demonstrate that ~10-MHz radio attenuation associated with optical PsA and other auroral activity can be detected using the SYE SuperDARN radar. The decreased background noise and echo power caused by enhanced ionization in the *D* region can be readily identified in quick-look plots similar to Figures 2 and 3. We now use the radar data to estimate the start and end times of a large set of PsA events initially identified from the optical data. We then determine and compare the durations and MLT distribution of PsA events identified from the camera and radar data sets.

A list of optical PsA events was generated for the imaging seasons 2011–2015. PsA were identified by visually inspecting daily keogram plots for patchy, low-intensity aurora. Since we intend to compare the camera observations with the radar data, intervals for which radar data were not available were excluded from the optical event list. No radar data were available for the 2012 imaging season, so the PsA event list covers the 2011 and 2013–2015 seasons. With these selection criteria, a total of 119 PsA events was identified. This event list is not exhaustive since the camera observations are not possible during periods of daylight or cloud cover. Also, it is common for PsA to drift outside of the camera FOV, in which case they pertain to a different geographic location and represent the end of the event at Syowa station. Therefore, the event start and end times relate to periods when the PsA is the dominant auroral feature in the camera FOV.

For each PsA event, quick-look plots similar to Figures 2 and 3 were examined for evidence of radio attenuation, and the start and end times recorded. Event onset times for the radar data set were identified when there was a sudden decrease in the background noise accompanied by backscatter loss on multiple radar beams. Similarly, event end times were determined based on the sudden return of *F* region backscatter and a simultaneous increase in the background noise. As demonstrated in section 3, the radar responds rapidly to the presence and absence of PsA and other EEP events, so EEP-related attenuation can be distinguished from the more gradual diurnal variation in the background noise in almost all cases. Fifteen events from the original PsA event list were rejected because there was no clear evidence of any attenuation in the radar data. These events were mostly very low intensity PsA or events which were observed only in the equatorward portion of the camera FOV. A further two events were rejected because the end time could not be clearly identified in the radar data, perhaps due to a more gradual tapering of the particle flux, which produces a gradual rather than sudden increase in the background noise. The final event list contained 102 events for which both an optical and radar signature were observed. The mean (median) event duration determined from the camera data was 2.04 hr (1.67 hr), compared to 3.34 hr (2.83 hr) determined from the radar data. The longest event identified in the camera data lasted for 5 hr 50 min and was still in progress at the end of imaging. Events lasting up to 9 hr were identified in the radar data.

Figure 4a shows a scatter plot comparing the camera- and radar-determined start times of the PsA events. The red symbols indicate events for which the camera-determined start time is uncertain due to cloud cover (10 events), and the black symbols indicate optical PsA with known start times (92 events). There is good agreement between the camera- and radar-determined start times, with most events lying close to the line of equality (dashed line). In the premidnight sector, radar-determined start times are 30–60 min earlier than camera-determined start times. This occurs due to the substorm activity preceding the PsA, which produces a response in the radar data before PsA are clearly identifiable in the camera data.

Figure 4b shows a scatter plot comparing the camera- and radar-determined event end times, in the same format as Figure 4a. Out of 102 PsA events in the list, the end time could be reliably determined from the camera data for only 33 events. The end of imaging was the most common reason that the event end time could not be determined from the camera data (64 events). For the 33 events with known camera-determined end times, there is good agreement between the radar and camera observations. After ~04:00 MLT, the event end times can no longer be determined from the camera data due to daylight, but the radar observations show that the EEP continues several hours into the morning sector. Note that after the camera stops imaging, we cannot verify that the observed radio attenuation is indeed a response to PsA. Therefore, the events used to produce Figure 4b correspond to continuous periods of HF attenuation which *commenced* during a period of optically confirmed PsA.

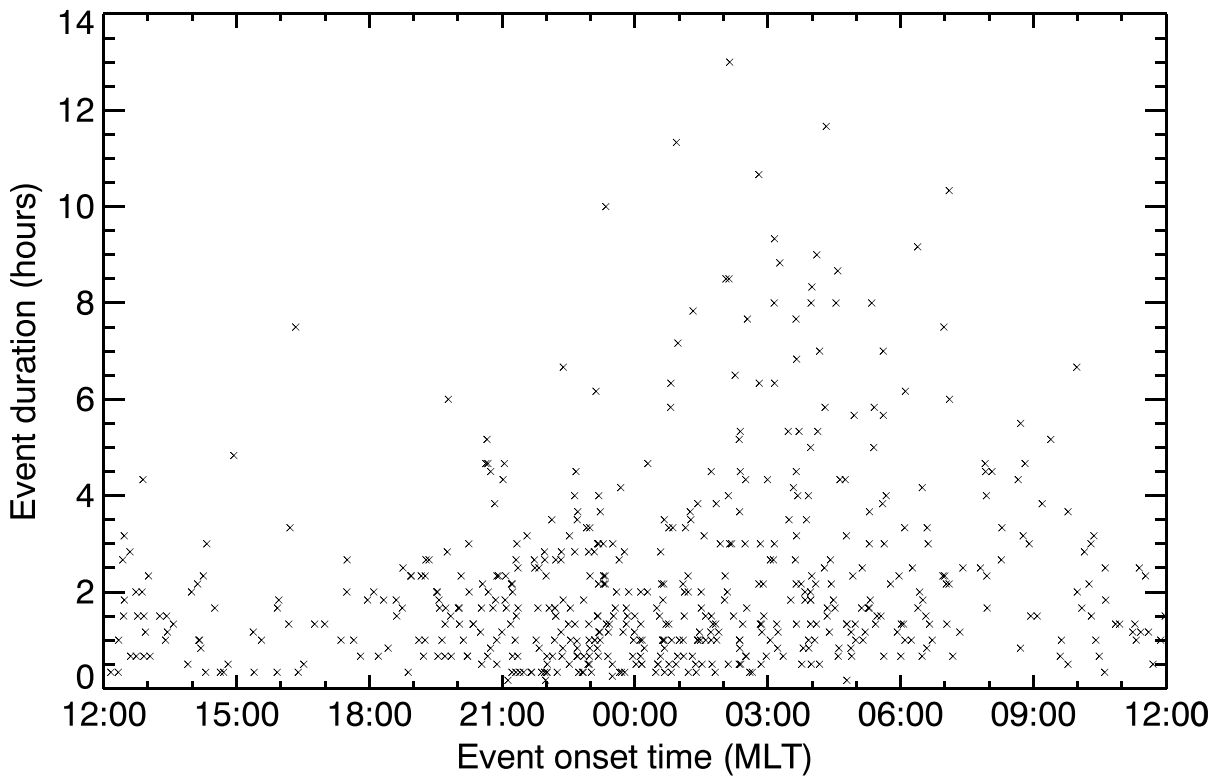

**Figure 6.** Duration of energetic electron precipitation events versus event onset time. MLT = magnetic local time.

The MLT distribution of the PsA events is shown in Figure 5. The solid red (black) line shows the MLT distribution determined from the optical (radar) data. The dashed lines show the MLT distributions of events, which occurred in June and July when the camera is able to detect events further into the morning sector. To produce this plot, we created 10-min bins of MLT and counted the number of times in which the camera/radar detected an event in each bin. We then averaged the results into hourly MLT bins. The area under each histogram curve equals the combined event duration in MLT hours. Since we have used the optical data as a starting point for identifying the PsA signatures in the radar data, it is expected that the two data sets will have similar MLT distributions. The agreement between the camera and radar observations is particularly good between 18:00 and 02:00 MLT. After 02:00 MLT, the camera-determined occurrence decreases rapidly due to daylight, whereas the radar continues to observe attenuation signatures for a further 2–4 hr. This result is even clearer if we consider only the events surrounding the winter solstice (dashed lines), where the event occurrence is almost identical for both the camera and radar data sets before 04:00 MLT, but then the camera-determined PsA occurrence falls sharply when optical observations are no longer possible. Although we cannot prove that the PsA is responsible for the attenuation measured by the radar after the camera stops imaging, the rapid dropoff in optical PsA occurrence at 04:00 MLT suggests that PsA-related EEP does continue further into the morning sector than is observable with the camera. Note that the zero occurrence from 11:00–18:00 MLT is related to the absence of observations (no camera data) rather than the absence of EEP.

## 5. Radar-Derived EEP Occurrence Rates

The radar data set makes it possible to produce a more complete event list, which is not restricted by cloud cover or daylight. Since PsA are associated with low particle fluxes into the *D* region (and hence low attenuation), our event detection method will also identify other EEP events such as substorms, as demonstrated in section 3. Without the optical data to verify that the attenuation signatures observed in the radar data are related to PsA, the new event list comprises all radio attenuation events resulting from the ionization of the lower thermosphere by any EEP source.

To produce our EEP event list, we generated 24-hr quick-look plots similar to Figures 2 and 3 for the whole of 2011 and visually identified radio attenuation events using the background noise and echo power parameters. Event start times were identified as a sharp decrease in both of these parameters across most of the FOV, as shown in Figures 2 and 3. Similarly, end times were defined as a rapid increase in background noise and the return of *F* region backscatter. Care was taken to ensure that discontinuities in the radar data due to changes in the radar operational mode were not erroneously identified as EEP events. Such discontinuities may arise from changes to the radar operating frequency, range gate separation, or integration time, for example. Five hundred fifty-five individual events were identified over the 12-month interval, with a combined duration of 1,251 hr. The individual event durations are plotted as a function of their onset time in Figure 6. Event onsets are observed at all MLT hours, but most occur on the nightside (18:00–06:00 MLT). Event durations ranged from 10 min to 13 hr, with a mean duration of 2.25 hr (median 1.67 hr). Ten percent of the events had durations of more than 5 hr, and almost all of these longer-duration events occur in the postmidnight/early morning sector.

The EEP event list was then used to estimate EEP occurrence rates as a function of MLT. Occurrence rates were determined in 1-hr MLT bins and represent the percentage probability of observing EEP in that MLT bin. These occurrence rates were calculated relative to the total radar observation time in each MLT bin. The total observation time per month ranged from 20–31 days for each MLT bin. The total observation time was determined as the amount of time in which the radar operated in the 10.0- to 10.6-MHz frequency range and sampled all 16 beams every minute. Intervals with strong radio interference (very high noise levels and no grouped populations of echo returns) were excluded. The seven solar proton events that occurred in 2011 were also excluded because the polar cap absorption is too strong for any less intense attenuation to be detectable (total 14 days).

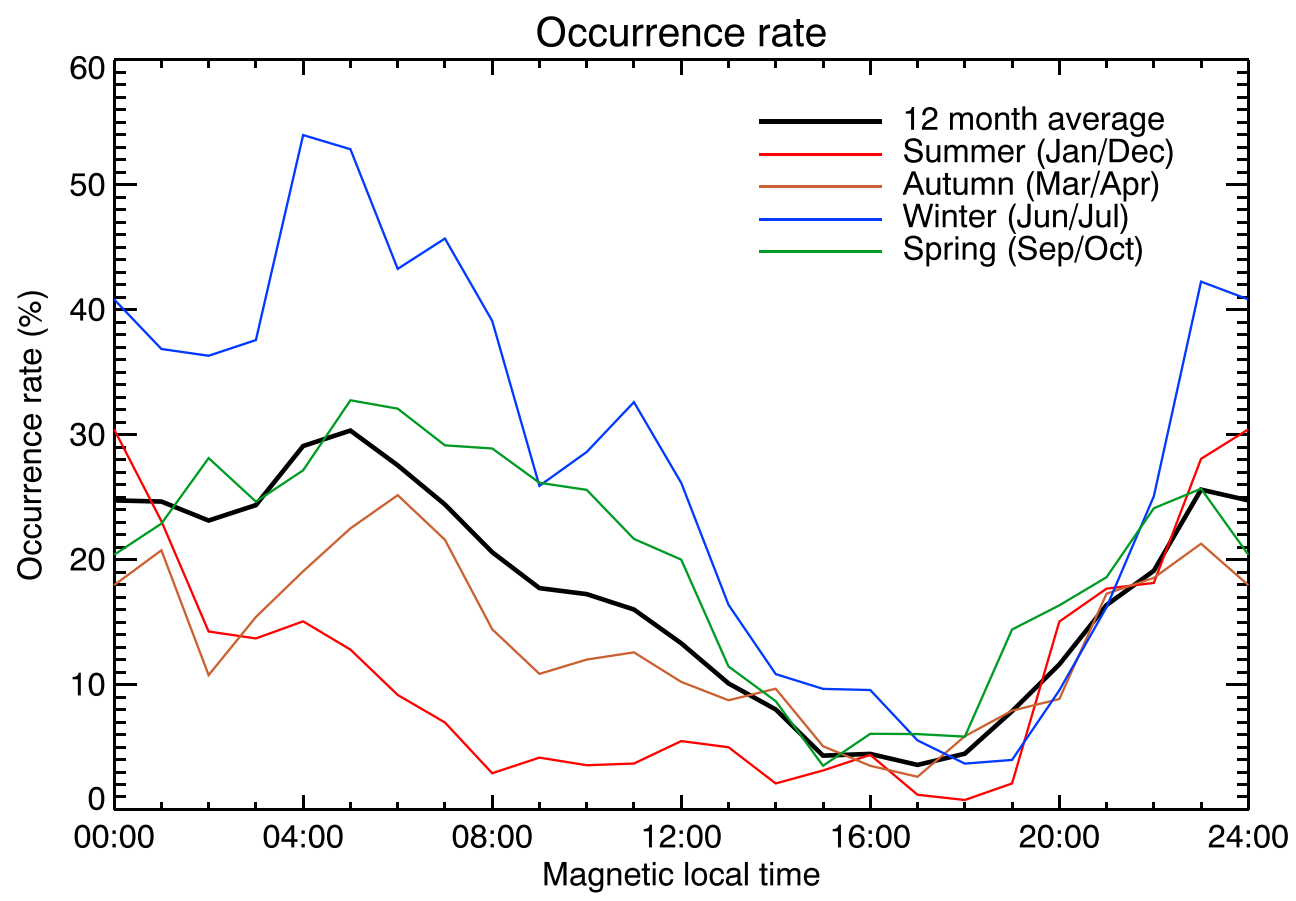

**Figure 7.** Energetic electron precipitation occurrence rates for 2011 as a function of magnetic local time. The black line shows the 12-month average occurrence rates, and the colored lines show occurrence rates for the two calendar months surrounding the solstices and equinoxes.

EEP occurrence rates determined from the 2011 radar data set are shown in Figure 7. The black line shows the 12-month average occurrence rate, and the colored lines show seasonal occurrence rates determined from the two calendar months surrounding the solstice/equinox in each season. The 12-month average occurrence rates have a similar MLT dependence to the earlier results shown in Figure 5, with a maximum of about 30% at ~04:00 MLT. In the afternoon sector, the EEP occurrence rate reaches a minimum (~5%) before increasing to about 25% near magnetic midnight. There are, however, significant seasonal differences in the EEP occurrence rates. Occurrence rates in the postmidnight sector were highest in the winter (June and July) and lowest in the summer (January and December). For the equinoctial months, postmidnight sector occurrence rates are in the ~20–30% range. In the afternoon and evening sectors, occurrence rates are similar for all seasons.

## 6. Discussion

The SYE SuperDARN radar has been used to detect HF radio attenuation in the *D* region ionosphere due to ionization by precipitating energetic electrons (tens to hundreds of kiloelectron volts) and hence determine EEP occurrence rates over Syowa station. HF radio attenuation can be identified in the SYE radar data as a reduction in both the echo power and background noise parameters. The data show that EEP ionization as weak as PsA produces a clear attenuation signature in the radar data. Therefore, EEP events involving higher fluxes of energetic electrons such as substorms are also detectable using this method.

The first part of this study (section 4) focused specifically on PsA-related EEP. Starting from a list PsA events identified using the all-sky camera data, we used the radar data to determine the start and end times of the PsA-related EEP. In general, PsA start times determined from the camera and radar data were similar, indicating that the onset of optical PsA coincides with increased ionization in the *D* region (Figure 4). In the premidnight sector, all camera-determined onset times lag behind the radar-determined onset times by up to 1 hr. This time offset occurs because PsA are often obscured by other auroral features in the keogram data, especially substorm-related aurorae (Partamies et al., 2017). For example, the first optical PsA event in Figure 2 emerges from the brighter auroral activation which ends at 00:40 UT. While we cannot determine whether the optical PsA have already commenced while the brighter, substorm-related aurorae are still in progress, it is clear from the radar data that the EEP ionization commences before isolated PsA are visible in the camera data. The noise level increases gradually following the transition from non-PsA to PsA (Figure 2b), indicating that the radar detects attenuation in response to the substorm-related aurora and then continues to respond to the weaker electron flux during the PsA.

Differences in the camera- and radar-determined start/end times may also arise due to differences in the FOVs of the two instruments. For the radar data, we expect that the primary attenuating region is very close to the radar site, in the area shaded blue in Figure 1. We draw this conclusion based on two observations. First, most backscatter detected by the SYE radar is direct (half-hop) ionospheric backscatter, so the transmitted waves traverse the *D* region only in the near-range FOV. Second, the background noise measured across the radar FOV during PsA events closely follows the location of the optical aurora. For example, in Figure 3, the noise levels measured along Beams 14 and 15 are lower than for other beams when the optical aurora shifts equatorward at 23:50–00:20, 02:40–03:20, and 04:00–04:30 UT. This effect is also apparent in Figure 2 at ~02:40–05:00 UT. This demonstrates that the radar is also sensitive to spatial variations in the EEP flux, and since the near-range radar FOV overlaps with only a small portion of the camera FOV, differences of up to ~1 hr in event onset/end times are to be expected.

In the second part of this study (section 5), we used a full year of radar data to determine EEP event durations and occurrence rates at Syowa station. In the absence of camera data, the presence of optical PsA can only be speculated based on the results of section 4, so we shifted the focus to detecting any type of EEP event.

Earlier studies using all-sky cameras have found that PsA occurrence rates peak in the morning sector and typically last for several hours (Jones et al., 2011, 2013; Partamies et al., 2017). By comparing the shape of the MLT distributions in Figures 5 and 7 and also noting that most long-duration ($\gtrsim$5 hr) EEP events occur in the postmidnight/early morning sectors (Figure 6), we can conclude that a significant fraction of the long-duration EEP events in the postmidnight/early morning sector are likely to be PsA. Short-duration events, especially those in the premidnight sector, are probably related to substorm activity (Frey et al., 2004).

It should be noted that some attenuation events lasting less than about 1.5 hr on the dayside may actually be shortwave fadeout events caused by solar X-ray flares and are therefore not related to EEP (Berngardt et al., 2018; Chakraborty et al., 2018). However, out of 30 solar flares detected by the Geostationary Operational Environmental Satellite (GOES-15) in 2011, we found only three instances in which a flare coincided with an attenuation event at SYE during sunlit conditions (not shown). Since shortwave fadeout events are short-lived and affect only the sunlit ionosphere, they are unlikely to be a significant contaminant to the EEP occurrence rates shown in Figure 7. The EEP occurrence rates are also unlikely to be contaminated by protons because proton precipitation occurs almost exclusively during solar proton events, which were explicitly excluded from the radar data set.

A significant advantage of the radar data over optical observations is that the EEP occurrence rates could be determined for all MLTs, seasons, and terrestrial weather conditions (Figure 7). In the 16:00–22:00 MLT sector, EEP occurrence rates are similar for all four seasons. At other times, however, the occurrence rates exhibit significant seasonal variations, reaching a maximum in the winter (June/July) and minimum in the summer (January and December). The postmidnight/morning sector wintertime EEP occurrence rates found here (50–55%) agree well with optical PsA occurrence rates determined by Jones et al. (2011; 55–65%), which suggests that most morning sector EEP events are related to PsA. Morning sector EEP occurrence rates are around 20–30% in the equinoctial months, which is close to the 12-month average. Around the summer solstice (January and December), EEP occurrence peaks at magnetic midnight and there is no discernible peak in the morning sector. It is not clear whether the seasonal variations reported here are a unique feature of the 2011 data set or if the results are repeatable for other years. Nightside auroral electron precipitation is known to have higher number fluxes and higher average energies in the winter hemisphere (Liou et al., 2001; Newell et al., 2010). In particular, in the postmidnight sector at the equatorward edge of the auroral oval, the average energy of precipitating electrons in the winter is almost double that of the summer (Liou et al., 2001). Although these studies include all auroral precipitation, the higher average energy of the nightside auroral electron precipitation in the winter is consistent with the higher wintertime EEP occurrence rates determined from the SYE radar data.

The seasonal and diurnal behavior reported in this study are also consistent with earlier statistical studies of CNA using riometers. Multiple riometer studies report lower levels of CNA in the morning sector during the summer, and no seasonal differences in the afternoon and evening sectors (e.g., Basler, 1963; Foppiano & Bradley, 1985; Kavanagh, 2002; Kavanagh et al., 2012). While these studies consider mean CNA rather than event occurrence, it is reasonable to compare our results to the earlier riometer work because we expect a correlation between event occurrence and mean CNA (Kavanagh, 2002). A key difference between our results and those from the riometer studies is the location of the daily maximum. The maximum CNA typically occurs at around 09:00–10:00 MLT, compared to the occurrence maximum at 04:00–05:00 MLT in our study. This might be related to the presence of weak PsA-related EEP in the earlier morning sector, which contributes significantly to the event occurrence rate but produces only small enhancements in the CNA.

Although the EEP occurrence rates shown in Figure 7 probably represent a real seasonal variation in the EEP energy input to the *D* region, there are some aspects of the radar-based event detection method, which might reduce the event detection efficiency in the summer. In particular, Syowa station is exposed to approximately 2 months of continuous daylight in the summer, resulting in increased radio attenuation due to photoionization of the *D* region. This non-EEP-related attenuation may make it difficult to identify the sudden backscatter loss at the onset of an EEP event. Furthermore, the radar-based EEP detection method requires that coherent backscatter is present immediately prior to EEP event onset so that the reduced echo power can be observed. If no coherent backscatter is present at EEP onset time, the event will not satisfy our selection criteria and the EEP occurrence rates will be underestimated. This is most likely to occur during the summer months when SuperDARN radars observe lower quantities of ionospheric backscatter (Ghezelbash et al., 2014; Koustov et al., 2004; Milan, Yeoman, et al., 1997; Ruohoniemi & Greenwald, 1997).

This study demonstrates that SuperDARN HF radars can be used to detect *D* region ionization caused by EEP, including low number flux EEP associated with PsA. The low operating frequency range (∼10–15 MHz) of most SuperDARN radars is advantageous for detecting EEP because lower radio frequencies are more strongly attenuated in the ionosphere than the higher frequencies used by riometers. Assuming an inverse square relationship between nondeviative absorption and frequency (e.g., Davies, 1990), the typical 1 dB of CNA measured by a 30-MHz riometer during a PsA event would correspond to 9-dB attenuation for a SuperDARN radar operating at 10 MHz. Due to the oblique orientation of the SuperDARN antenna radiation pattern, and the fact that half-hop ionospheric backscatter traverses the *D* region twice, the amount of attenuation would be even higher. This almost completely extinguishes the coherent backscatter at SYE during the EEP event, a feature which is readily identifiable in the quick-look plots. For some events, small amounts of near-range coherent backscatter persist throughout the event (e.g., 00:20–02:20 UT in Figure 2c). EEP may produce nonaspect sensitive irregularities in the lower *E* and *D* regions, which are readily detectable by SuperDARN radars (Milan et al., 2004, 2008), and since echo power attenuates with range as approximately $1/r^3$, it is possible for long-range (*F* region) coherent scatter to be lost while short-range scatter is still discernible above the noise level. For the EEP event at 00:20–02:20 UT in Figure 2, the power of the near-range backscatter decreases sharply at event onset, indicating that the signals transmitted by the radar were indeed attenuated in the *D* region during the EEP event but did not completely extinguish the near-range backscatter.

As of 2019, SuperDARN consists of 36 radars with near-FOVs covering subauroral, auroral, and polar latitudes, and the data set consists of more than 20 years of observations. The wide latitudinal coverage of SuperDARN would enable the radars to detect EEP events under a range of geomagnetic conditions. The near-range FOV coverage is particularly good at around 60–65° magnetic latitude, which includes the equatorward edge of the nightside auroral oval under low to moderate geomagnetic conditions. Therefore, many radars are well positioned to detect radio attenuation due to EEP. We anticipate that our EEP event detection method could be automated and then applied to a multiradar data set to determine EEP occurrence rates over a longer time period.

## 7. Conclusion

Observations of ∼10-MHz radio noise and echo power from the SYE SuperDARN radar have been used to detect radio attenuation in the *D* region ionosphere due to EEP events. By comparing the start/end times of PsA events determined from the radar and all-sky camera data, we conclude that HF attenuation that commences during periods of optical PsA continues for typically 2–4 hr after the camera stops imaging at dawn. We then generated a database of 555 EEP events identified in the radar data during the year 2011 and showed that EEP occurrence rates in the postmidnight/early morning sector exhibit significant seasonal variations. EEP occurrence rates are highest during the winter, reaching ∼50–55% at 04:00–05:00 MLT. Occurrence rates in the equinoctial months peak at around 05:00–06:00 MLT but are lower overall (∼20–30%). In the summer, there is no distinct occurrence peak in the early morning sector, and instead, the EEP occurrence rate peaks at magnetic midnight (30%). We suggest that many other SuperDARN radars would be capable of detecting EEP-related radio attenuation, which may be useful in a larger study of EEP occurrence rates.

**Acknowledgments**
This study was supported by the Research Council of Norway/CoE under contract 223252/F50. The CDC (All-sky Color Digital Camera) at Syowa station and the SENSU Syowa East (and South) SuperDARN radars are part of the Science Program of Japanese Antarctic Research Expedition (JARE) and are supported by NIPR under the Ministry of Education, Culture, Sports, Science and Technology (MEXT), Japan. The SuperDARN data were obtained from the British Antarctic Survey data mirror (https://www.bas.ac.uk/project/superdarn). The CDC data are available at the http://polaris.nipr.ac.jp/~acaurora/aurora/Syowa/ website. The solar flare event list was obtained from the Space Weather Database Of Notifications, Knowledge, Information (DONKI, https://kauai.ccmc.gsfc.nasa.gov/DONKI/search/).